\documentclass[a4paper,fleqn]{cas-dc}

\usepackage[numbers]{natbib}

\def\tsc#1{\csdef{#1}{\textsc{\lowercase{#1}}\xspace}}
\tsc{WGM}
\tsc{QE}

\begin{document}
\let\WriteBookmarks\relax
\def\floatpagepagefraction{1}
\def\textpagefraction{.001}

\shorttitle{Energy-Guided Diffusion Model for CBCT-to-CT Synthesis}    

\shortauthors{Linjie Fu et al.}

\title[mode = title]{Energy-Guided Diffusion Model for CBCT-to-CT Synthesis}

\author[1,2]{Linjie Fu}[
    orcid=0000-0002-9109-2670]
\ead{fulinjie19@mails.ucas.ac.cn} 

\author[4]{Xia Li}
\ead{lixia_rt_wch@scu.edu.cn}

\author[1,2]{Xiuding Cai}
\ead{caixiuding20@mails.ucas.ac.cn} 

\author[1,2]{Dong Miao}
\ead{miaodong20@mails.ucas.ac.cn} 

\author[1,2]{Yu Yao}
\ead{casitmed2022@163.com} 

\author[3]{Yali Shen}
\ead{sylprecious123@163.com} 

\address[1]{Chengdu Computer Application Institute Chinese Academy of Sciences, China}
\address[2]{University of the Chinese Academy of Sciences, China}
\address[3]{Sichuan University West China Hospital Department of Abdominal Oncology, China}
\address[4]{Radiophysical Technology Center, Cancer Center, West China Hospital, Sichuan University, China}

\begin{abstract}
Cone Beam CT (CBCT) plays a crucial role in Adaptive Radiation Therapy (ART) by accurately providing radiation treatment when organ anatomy changes occur. However, CBCT images suffer from scatter noise and artifacts, making relying solely on CBCT for precise dose calculation and accurate tissue localization challenging. Therefore, there is a need to improve CBCT image quality and Hounsfield Unit (HU) accuracy while preserving anatomical structures. To enhance the role and application value of CBCT in ART, we propose an energy-guided diffusion model (EGDiff) and conduct experiments on a chest tumor dataset to generate synthetic CT (sCT) from CBCT. The experimental results demonstrate impressive performance with an average absolute error of 26.87±6.14 HU, a structural similarity index measurement of 0.850±0.03, a peak signal-to-noise ratio of the sCT of 19.83±1.39 dB, and a normalized cross-correlation of the sCT of 0.874±0.04. These results indicate that our method outperforms state-of-the-art unsupervised synthesis methods in accuracy and visual quality, producing superior sCT images.
\end{abstract}




\maketitle

\section{Introduction}
In the field of medicine, Adaptive Radiation Therapy (ART) is an advanced treatment modality that involves real-time and dynamic adjustments to the treatment plan by monitoring and managing biological and physiological changes in patients during the radiation therapy process through the use of Cone-beam CT (CBCT)\cite{Yan_Vicini_Wong_Martinez_1997,Lim-Reinders_Keller_Al-Ward_Sahgal_Kim_2017}. Continuously adjusting the radiation therapy plan throughout the entire treatment process enhances the precision of the treatment and improves the quality of life for patients.
As a standard daily imaging modality for radiation therapy patients, the effective use of Cone Beam CT (CBCT) in the context of Adaptive Radiation Therapy (ART) has been extensively studied\cite{Posiewnik_Piotrowski_2019,Wu_Li_Wu_Yin_2011,barateau2020comparison}. However, it faces some challenges in the implementation. CBCT typically provides lower image quality compared to the CT images used for initial treatment planning due to the presence of scatter noise and artifacts. Consequently, relying solely on CBCT images for accurate dose calculation and precise tissue localization is not feasible\cite{Kim_Yoo_Yin_Samei_Yoshizumi_2010}. Therefore, accurately converting real-time CBCT images into planned CT images for precise radiation therapy is an urgent and critical scientific problem that needs to be addressed.

Physics-based approaches such as analytical modeling methods\cite{Naimuddin_Hasegawa_Mistretta_1987,Veldkamp_Thijssen_Karssemeijer_2003}, advanced iterative reconstruction methods\cite{Sidky_Kao_Pan_2006,Tian_Jia_Yuan_Pan_Jiang_2011}, Monte Carlo simulations\cite{Mainegra-Hing_Kawrakow_2010}, rule-based methods with prior knowledge\cite{Zhang_Yin_Segars_Ren_2013}, and random forests\cite{Wang_Gao_Shi_Li_Chen_Tang_Xia_Shen_2015} have been applied to CBCT-to-CT conversion. However, the accuracy in handling complex biological tissue density distributions still needs improvement. Recently, deep learning has demonstrated significant potential and remarkable success in medical image analysis tasks such as denoising\cite{QingsongYang2018LowDoseCI,HuChen2017LowDoseCW}, super-resolution\cite{QingLyu2019MRISW,KerstinHammernik2017LearningAV}, and artifact reduction\cite{YanboZhang2017ConvolutionalNN,LarsGjesteby2017DeepLM}. Compared to traditional methods, deep neural networks learn features in a data-driven manner and generate superior feature representations. Therefore, GAN-based approaches\cite{DongNie2016EstimatingCI,XiaoHan2017MRbasedSC,AgisilaosChartsias2017AdversarialIS,YutaHiasa2018CrossModalityIS,LeiBi2017SynthesisOP,KarimArmanious2020MedGANMI} utilizing cycle consistency loss and contrastive loss have become the primary solutions for CBCT-to-CT synthesis. However, the biggest challenge faced by these methods is the simultaneous adversarial training of the generator and discriminator networks, which can limit the quality and diversity of synthesized images due to early convergence and mode collapse.

In recent years, diffusion models have achieved great success in the field of computer vision, with many studies successfully applying them to unsupervised generation tasks, resulting in high-fidelity natural images\cite{YangSong2019GenerativeMB,JonathanHo2020DenoisingDP,YangSong2021ScoreBasedGM}. Diffusion models are a powerful class of generative models. They start with observed data and transform it into latent standard normal noise by simulating a diffusion process, which is modeled as a Markov chain. Each step of this Markov chain slightly alters the data, progressively bringing it closer to the latent noise. Once this process is learned, new data (such as images) can be generated by sampling random noise and reversing the diffusion process, using a series of inverse steps to transform the noise back into the target data space. Compared to GANs, diffusion models allow exposure to data with various levels of noise during the training process, leading to more generalized model training. Research has shown that diffusion models outperform GANs in several image generation tasks\cite{FlorinelAlinCroitoru2022DiffusionMI,PrafullaDhariwal2021DiffusionMB}.

In this study, we propose EGDiff, which utilizes the diffusion model DDPM for CBCT-to-CT synthesis. We design an energy-guided function that enables training in both the CBCT and CT domains, learning domain-specific and domain-independent features. In the reverse diffusion process, we retain domain-independent features and discard domain-specific features, improving the realism and reliability of synthetic CT (sCT) images. When comparing the images generated using EGDiff with those generated using GANs based on cycle consistency loss and contrastive loss, the results show that the sCT images generated by EGDiff are closer to real CT images and outperform other methods in various evaluation metrics.
The goal of this research is to achieve CBCT-to-CT image conversion in a more efficient and high-quality manner. We envision that this technology will contribute to achieving more accurate tumor localization and dose calculation in practical applications such as Adaptive Radiation Therapy (ART). 

Our key contributions are as follows:

1.We apply a diffusion model to non-paired image translation for improving the quality of CBCT images and extensively compare it with other models.

2.We design an energy-guided function to enhance the quality of CBCT images to meet the requirements of adaptive radiation therapy.

\section{Related work}
\subsection{GAN-based image synthesis method}
The most common image synthesis tasks are GAN methods based on cycle consistency loss and GAN methods based on contrast loss. \cite{Zhang_Li_Dai_Zhong_Wang_Yang} propose a multi-modal unsupervised representation disentanglement (MURD) learning framework, which separates the content, style, and artifacts representation from CBCT and CT images in the latent space for synthesizing CT images. \cite{ Deng_Hu_Wang_Huang_Yang_2022} introduce Respatch-CycleGAN to transform CBCT into synthetic CT images, reduce the presence of metal artifacts, and restore HU values to align with planning CT. \cite{dar2019image} performs multi-contrast MRI synthesis using a conditional GAN with pixel-level and cycle consistency loss functions.The study by \cite{Deng_Ji_Huang_Yang_Wang} incorporates an auxiliary branch with Diversity Block Branch (DBB) modules into the generator of CycleGAN to improve the quality of synthesizing Computed Tomography (sCT) from Cone Beam Computed Tomography (CBCT). \cite{Kang_Shin_Yang_Kim_Huh_Lee_Heo_Yi_O} develops a GAN model based on patchwise contrastive learning to generate sCT images from CBCT images. \cite{TaesungPark2020ContrastiveLF} is achieved by a contrast learning method for unpaired image transformation and reduces the network parameters by using only one set of generator and discriminator. \cite{ZiliYi2017DualGANUD} maximizes mutual information by learning the correspondence between input and output image patches using separate embeddings. \cite{WeilunWang2021InstanceWiseHN} proposes a new negative generator to mine hard negative examples for contrast learning in unpaired image-to-image translation. 
 \cite{FangnengZhan2023ModulatedCF} proposed MoNCE loss which uses optimal transmission\cite{GabrielPeyr2019ComputationalOT} to coordinate the re-weighting of all negative samples among multiple targets. \cite{XueqiHu2023QSAttnQA} forms a better feature suitable for contrast learning by selecting image-related anchor points as queries to obtain parts at other locations.

\subsection{Image synthesis method based on diffusion model}
In recent years, diffusion models have splashed on computer vision tasks, especially image synthesis tasks. The diffusion model performs T times add Gaussian noise on the image in the forward process and noise sampling in the reverse process to achieve image reduction. Given this potential, diffusion model recently use for unimodal imaging tasks such as image reconstruction\cite{AjilJalal2021RobustCS,HyungjinChung2023ScorebasedDM,SalmanUHDar2022AdaptiveDP,HyungjinChung2023MRID}, unconditional image synthesis\cite{WalterHLPinaya2022BrainIG}, and anomaly detection\cite{wolleb2022diffusion,WalterHLPinaya2023FastUB}. \cite{peng2023cbct} designs a conditional denoising diffusion probabilistic model (DDPM) for generating sCT from CBCT, the results demonstrate that this method achieves excellent outcomes in terms of both visual quality and quantitative analysis. \cite{WalterHLPinaya2022BrainIG} used diffusion models to create synthetic MRI images of the adult brain but did not achieve image to image synthesis. \cite{QingLyu2022ConversionBC} proposes diffusion and score-matching models for image synthesis between CT and MRI. \cite{XieTaofeng2022BrainPS} Using Joint Probability Distribution of Diffusion Model (JPDDM) for MRI-oriented synthesis of PET. \cite{WeiPeng2022GeneratingR3} train a Conditional Diffusion Probabilistic Model for 3D MRI synthesis. \cite{ZihaoWang2023ZeroshotLearningCD} proposes a new unsupervised zero-shot learning method, the Mutual Information Guided Diffusion Cross-Modality Data Translation Model (MIDiffusion).

Based on the above research, we propose EGDiff, which designs an energy-guided function for learning domain-independent and domain-specific features, and combines a diffusion model and a generative adversarial network to implement medical image synthesis. We have experimented with EGDiff on a chest tumor dataset and demonstrated that our approach could achieve excellent results in CBCT to CT image synthesis.

\section{Method}
\begin{figure*}[htbp]
\centering
\includegraphics[width=1\linewidth]{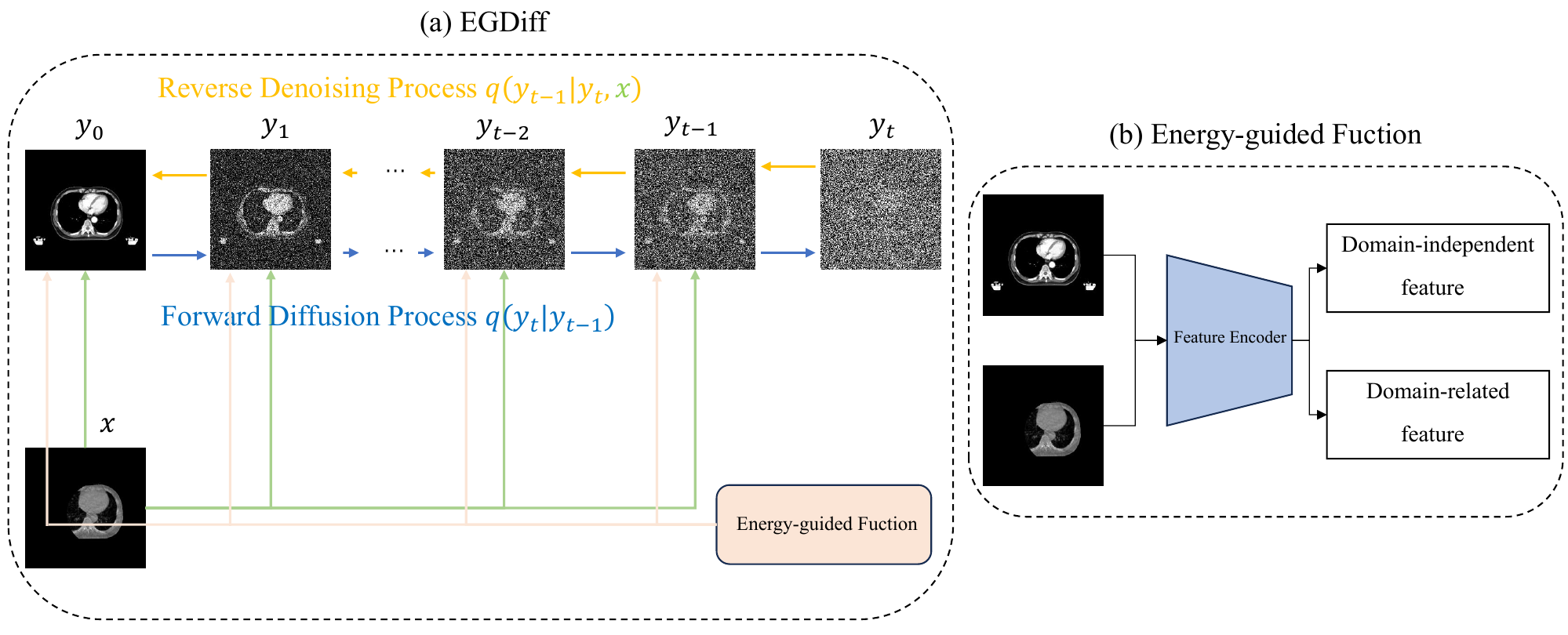}
   \caption{Workflows of (a) the framework of EGDiff,and (b) the proposed energy-guided function.}
\label{fig6}
\end{figure*}
Figure~\ref{fig6} gives the overall flow of our method. (a) shows the training framework of EGDiff, and (b) shows the training flow of the energy-guided function. In this section,  first we introduce the architecture of the conditional diffusion model, and we present the operation mechanism of the energy-guided function.
\subsection{Conditional Diffusion Model}
$x_{0}$ is the CBCT image, $y_{0}$ is the CT image, and our objective is to utilize a diffusion model to generate a synthesized CBCT image $x_{0}$ from the given CT image $y_{0}$. The Diffusion Denoising Probability Model (DDPM) employs Markov chain sampling and data generation to progressively transform random noise into the correlation between CBCT and CT images.
    
Given a CT image $y_{0}$ with step length T, the noise is first added gradually to $y_{0}$ by a forward diffusion process to obtain $y_{1}$, $y_{2}$,..., $y_{t}$. According to the \cite{JonathanHo2020DenoisingDP}, the noise added each time T obeys a Gaussian distribution $\mathcal N(0, I)$, so the distribution of $y_{t}$ obtained at step T can be expressed as the following Markov process:
\begin{equation}\label{eq1}
  q(y_{t}|y_{0})=\mathcal N(y_{t};\sqrt{\overline{\alpha}_{t}}y_{0},(1-\overline{\alpha}_{t})I)
\end{equation}
Where $\overline{\alpha}_{t}$:=$\prod_{n=1}^t\alpha_{n}$, $\alpha_{t}$:=$1-\beta_{t}$, $\beta_{t}$ is the forward process variances,  $I$ denotes the identity matrix. Ultimately, $y_{t}$ can be expressed as a function of $y_{0}$ as follows:
\begin{equation}\label{eq2}
  y_{t}=\sqrt{\overline{\alpha}_{t}}y_{0}+\sqrt{1-\overline{\alpha}_{t}}\varepsilon, \quad with\ \varepsilon \sim \mathcal N(0,I) 
\end{equation}
We try to denoise $y_{t}$ during the inverse diffusion and restore it to $y_{0}$. Since $q(y_{t-1}|y_{t})$ is unknown, we use the Bayesian formula by solving for:
\begin{equation}\label{eq3}
  q(y_{t-1}|y_{t},y_{0})= q(y_{t}|y_{t-1},y_{0})\frac{q(y_{t-1}|y_{0})}{q(y_{t}|y_{0})}
\end{equation}
According to Equation $(1)(3)$, the distribution of $q(y_{t-1}|y_{t})$ can be expressed as:
\begin{equation}\label{eq4}
  q(y_{t-1}|y_{t})= \mathcal N(y_{t-1};\overline{\mu}(y_{t},t),\overline{\beta} I)
\end{equation}
\begin{equation}\label{eq5}
  \overline{\mu}= \frac{\sqrt{\overline{\alpha}_{t-1}}\beta_{t}}{1-\overline{\alpha}_{t}}y_{0}+\frac{\sqrt{\alpha_{t}}(1-\overline{\alpha}_{t-1})}{1-\overline{\alpha}_{t}}y_{t}
\end{equation}
\begin{equation}\label{eq6}
  \overline{\beta}=\frac{1-\overline{\alpha}_{t-1}}{1-\overline{\alpha}_{t}}\beta_{t} 
\end{equation}
    According to Equation $(2)$, Equation $(5)$ can be rewritten as:
\begin{equation}\label{eq7}
  \mu_{\theta}(y_{t},t)=\frac{1}{\sqrt{\alpha_{t}}}(y_{t}-\frac{1-\alpha_{t}}{\sqrt{1-\overline{\alpha}_{t}}}\varepsilon_{\theta}(y_{t},t)) 
\end{equation}
In our methods, we use $x_{0}$ in the inverse diffusion process to guide $y_{t}$ for denoising, ultimately we can estimate $y_{t-1}$ from $y_{t}$ by:
\begin{equation}\label{eq8}
  y_{t-1}=\frac{1}{\sqrt{\alpha_{t}}}(y_{t}-\frac{1-\alpha_{t}}{\sqrt{1-\overline{\alpha}_{t}}}\varepsilon_{\theta}(y_{t},x_{0},t))+\sigma_{t}z  
\end{equation}
$z \sim \mathcal N(0,I)$, therefore, our objective is to train a neural network $\varepsilon_{\theta}$ based on the UNet architecture to estimate the noise at each step, and the \cite{JonathanHo2020DenoisingDP} gives the neural network optimization loss function, the loss function for training the network is defined as follows:

\begin{equation}\label{eq9}
\begin{split}
\mathcal{L}(\theta) = \mathbb{E}{y{t},x_{0},t}[|y_{t-1}-x_{0}|{2}^{2}]\\ + \lambda{t}\mathbb{E}{y{t},x_{0},t}[|\nabla_{y_{t-1}}\varepsilon_{\theta}(y_{t},x_{0},t)|_{2}^{2}]
\end{split}
\end{equation}

where $\mathbb{E}$ represents the expectation over the training data. The first term encourages the denoised image $y_{t-1}$ to be close to the ground truth CBCT image $x_{0}$. The second term penalizes the gradients of $\varepsilon_{\theta}$ with respect to $y_{t-1}$, which encourages smooth noise estimation.

In Equation $(9)$, $\lambda_{t}$ is a weighting factor that balances the contributions of the two terms. It can be chosen based on the specific requirements of the denoising task. Typically, $\lambda_{t}$ is set to increase over time $t$ to gradually emphasize the smoothness of the noise estimation.

By minimizing the loss function $\mathcal{L}(\theta)$ through backpropagation and gradient descent, the neural network $\varepsilon_{\theta}$ can be trained to estimate the noise during the inverse diffusion process, facilitating the generation of the synthesized CBCT image $x_{0}$ from the given CT image $y_{0}$.

\subsection{Energy-Guided Stochastic Differential Equations}
To fully utilize the information in both the target and source domains, we guide the inference process of the model using an energy function trained on both domains. For medical image synthesis, it is crucial to preserve domain-independent features (e.g., organ location) while modifying domain-related characteristics (e.g., morphological appearance). To achieve this, we define two potential logarithmic functions as energy functions that extract and aggregate domain-independent and domain-related characteristics, respectively. 

The domain adaptation potential function is defined using the cosine similarity between the original and source domain images, allowing us to discard domain-specific features and significantly enhance the realism of the synthesized image. It can be expressed as:

\begin{equation}\label{eq12}
S_a(y_t,x_0,t)=\frac{1}{HW}\sum_{h,w}\frac{F^{hw}(y_t,t)^\top F^{hw}(x_0,t)}{\Vert F^{hw}(y_t,t) \Vert_2\Vert F^{hw}(x_0,t) \Vert_2}  
\end{equation}

where $F^{hw}$ represents the channel features at spatial location $(h,w)$.

The domain discrepancy potential function aims to retain domain-independent features by calculating the negative squared L2 distance between the original and source domain images. It can be defined as:

\begin{equation}\label{eq13}
S_d(y_t,x_0,t)=-\Vert F^{hw}(x_0,t)-F^{hw}(y_{t},t) \Vert_2^2  
\end{equation}

Here, $F^{hw}$ is a domain-independent feature extractor represented by a low-pass filter.

Finally, the energy-guided function can be formulated as:

\begin{equation}\label{eq14}
S(y_t,x_0,t)=\lambda_{s}S_a(y_t,x_0,t)+\lambda_{d}S_d(y_t,x_0,t)  
\end{equation}

where $\lambda_{s}$ and $\lambda_{d}$ are weight hyperparameters representing the importance of the corresponding potential functions.

The total loss function in our method is defined as follows:

\begin{align}
\begin{split}
\mathcal{L}(\theta) = &\mathbb{E}_{y_t, x_0, t}[|y_{t-1}-x_{0}|^2_{2}] \\
&+ \lambda_{t_1}\mathbb{E}_{y_t, x_0, t}[|\nabla_{y_{t-1}}\varepsilon_{\theta}(y_{t},x_{0},t)|^2_{2}] \\
&+ \lambda_{t_2}\mathbb{E}_{y_t, x_0, t}[S(y_t, x_0, t)]
\end{split}
\end{align}

Here, $y_t$, $x_0$, and $t$ represent the target image, source domain image, and time step, respectively, and $\theta$ denotes the model's parameters.
\begin{figure*}[htbp]
\centering
\includegraphics[width=1\linewidth]{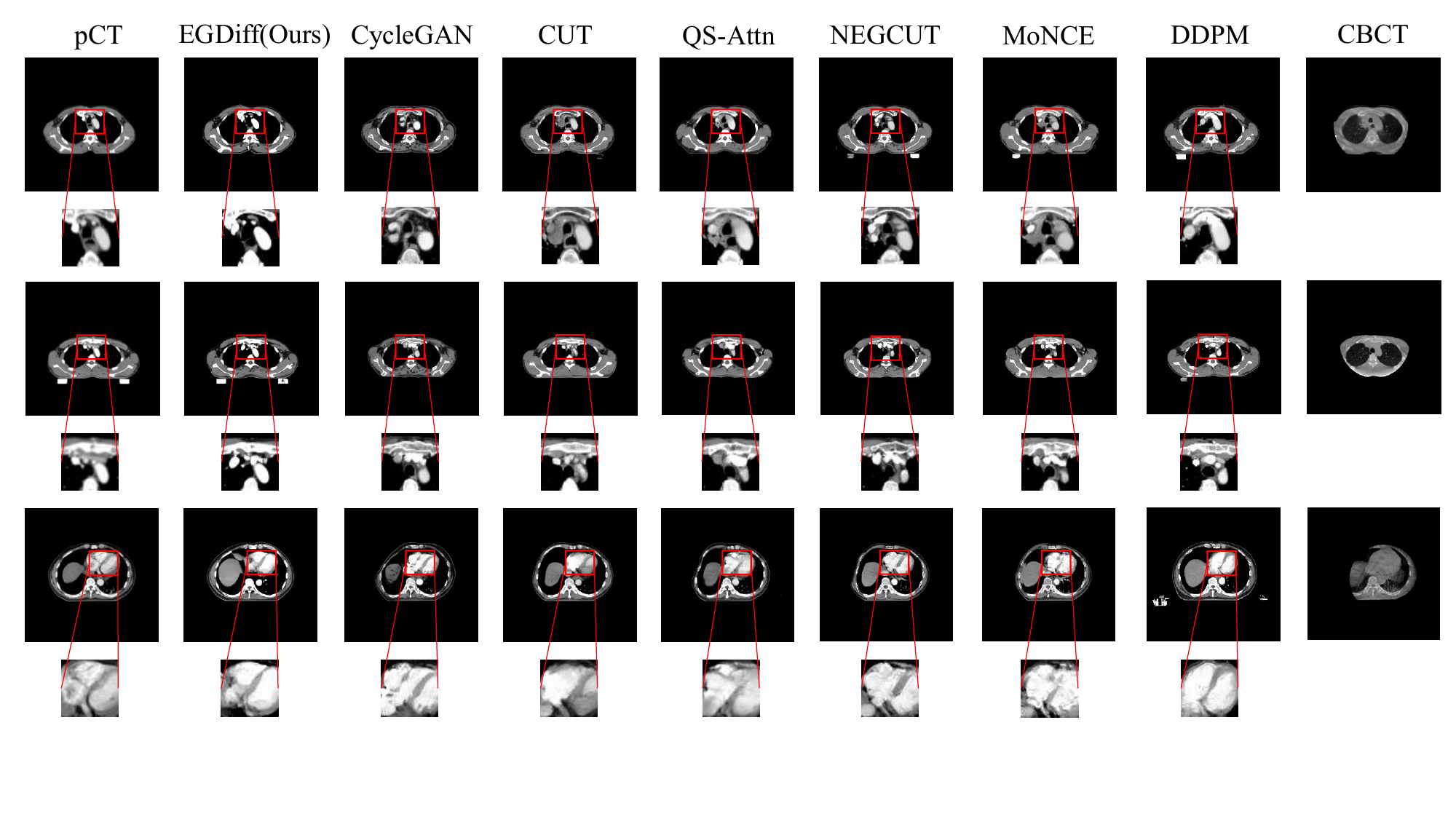}
   \caption{pCT, CBCT, and sCT generated by different models.The first, third, fifth row shows the image synthesis results, the second, fourth, sixth row shows the corresponding synthesis details.}
\label{fig2}
\end{figure*}

\begin{figure*}[htbp]
\centering
\includegraphics[width=1\linewidth]{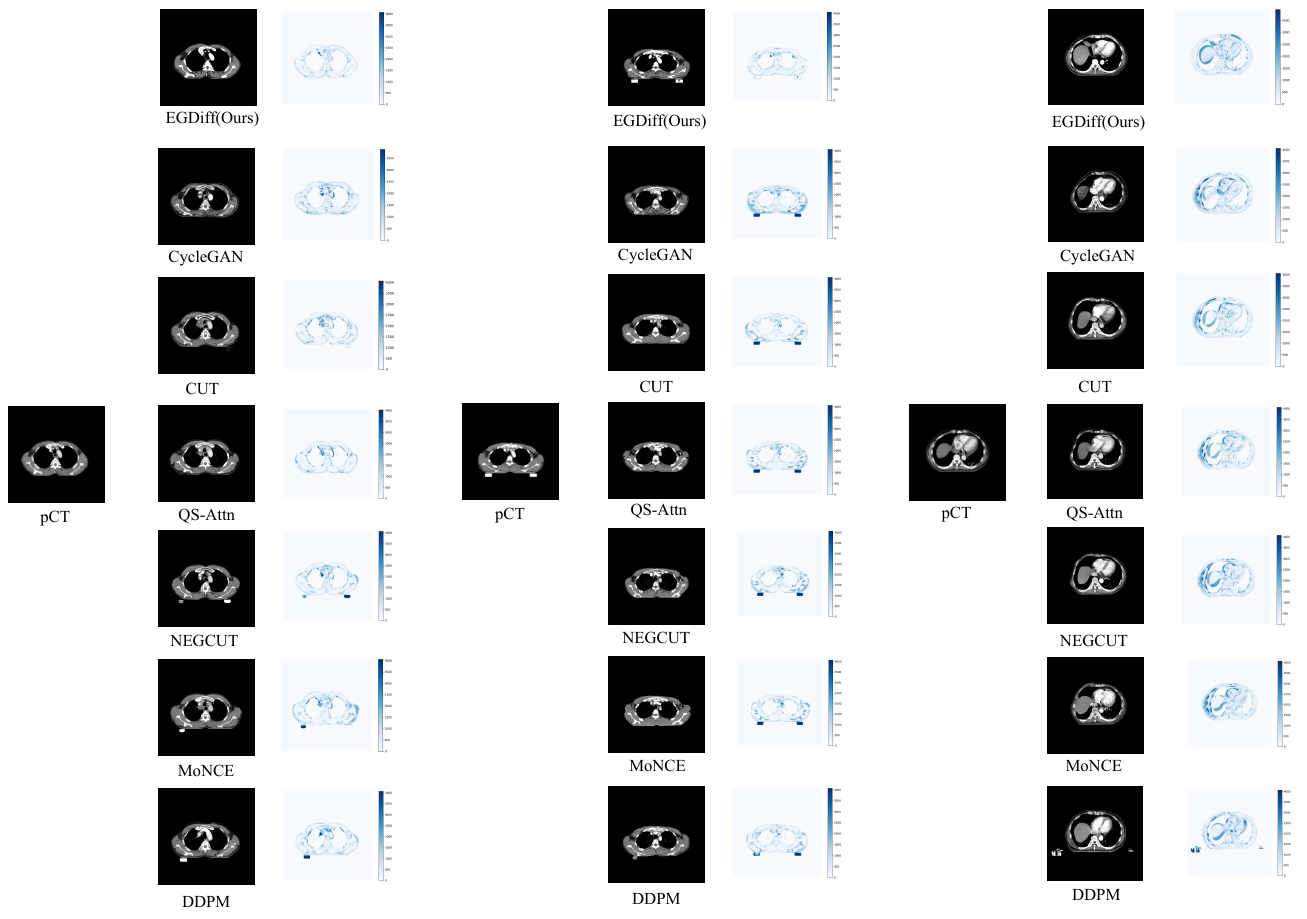}
   \caption{the HU value difference between the synthesized images and pCT.}
\label{fig1}
\end{figure*}
\begin{figure*}[htbp]
\centering
\includegraphics[width=1\linewidth]{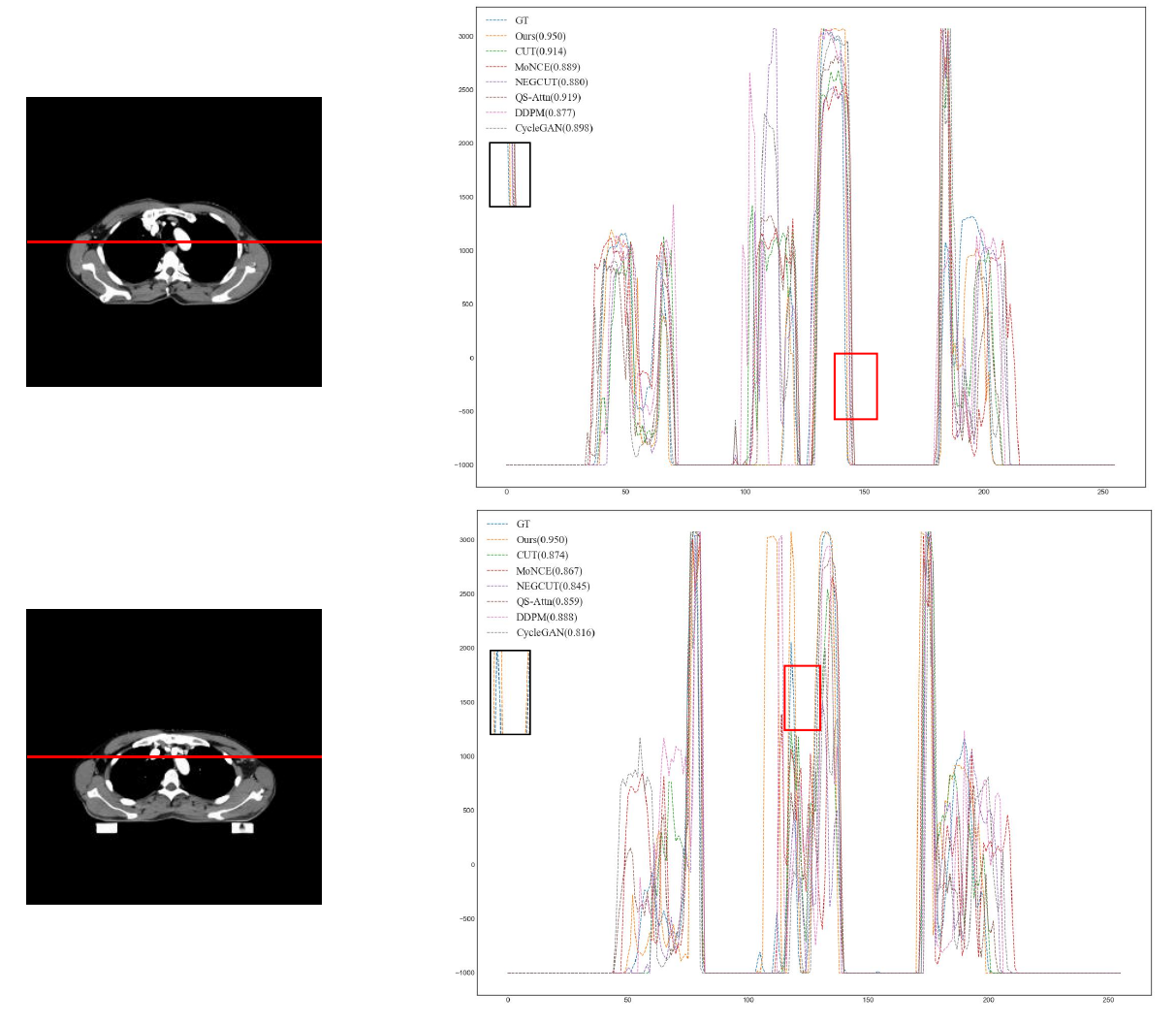}
   \caption{Horizontal Direction Hounsfield Unit (HU) Value Statistics, ranging from -1000 to 3071.}
\label{fig3}
\end{figure*}
\begin{figure*}[htbp]
\centering
\includegraphics[width=1\linewidth]{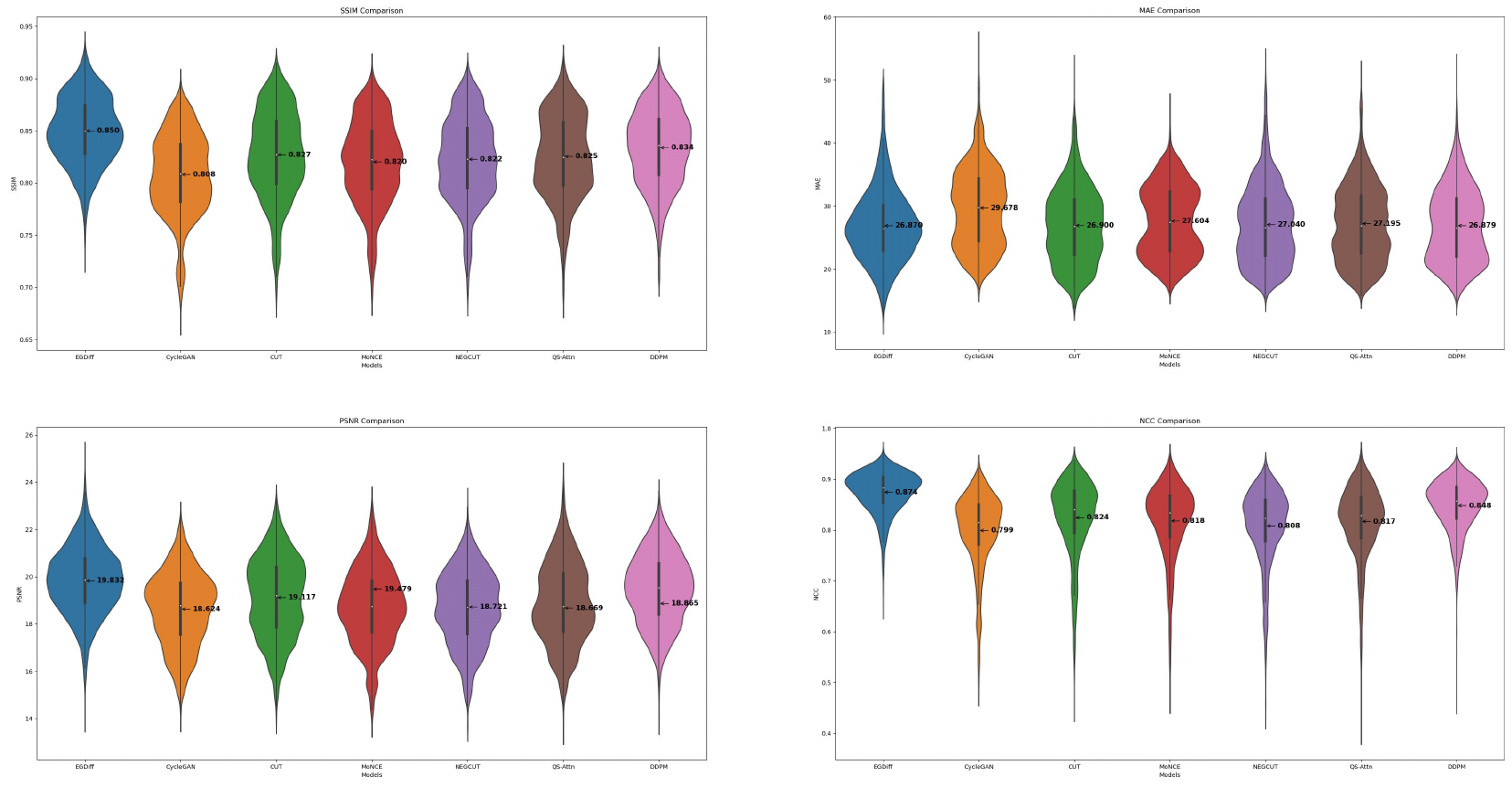}
   \caption{The Stats-Violin plot displays the distribution and average values of SSIM, MAE, PSNR, and NCC for seven different models synthesized sCT.}
\label{fig4}
\end{figure*}
\section{Experiment}
\subsection{Dataset}

\begin{center}
\begin{table*}[t]
\caption{Quantitative analysis results for assessing the quality of synthesized sCT.}
\label{table1}
\centering
\begin{tabular}{ccccc}
\Xhline{1.5 pt}
Method& SSIM$\uparrow$ & MAE(HU)$\downarrow$ & PSNR(dB)$\uparrow$ & NCC$\uparrow$ \\
\Xhline{0.5 pt}
EGDiff(Ours) &\pmb{0.850}$\pm$\pmb{0.03} & \pmb{26.87}$\pm$\pmb{5.90} & \pmb{19.83}$\pm$\pmb{1.39}& \pmb{0.874}$\pm$\pmb{0.04}\\
CycleGAN\cite{JunYanZhu2017UnpairedIT} &0.808$\pm$0.03 & 29.67$\pm$6.14 & 18.62$\pm$1.53& 0.799$\pm$0.07\\
CUT\cite{TaesungPark2020ContrastiveLF} &0.827$\pm$0.04 & 26.89$\pm$5.76 & 19.12$\pm$1.67& 0.824$\pm$0.08\\
NEGCUT\cite{WeilunWang2021InstanceWiseHN} &0.834$\pm$0.04 &26.88$\pm$5.82 & 19.48$\pm$1.48& 0.848$\pm$0.05\\
MoNCE\cite{FangnengZhan2023ModulatedCF} &0.820$\pm$0.04 &27.60$\pm$5.45 & 18.72$\pm$1.59& 0.818$\pm$0.07\\
Qs-Attn\cite{XueqiHu2023QSAttnQA} &0.822$\pm$0.04 & 27.04$\pm$6.10 & 18.67$\pm$1.53&0.808$\pm$0.07\\
DDPM\cite{Lyu_Wang_2022} &0.825$\pm$0.04 & 27.19$\pm$5.96 & 18.87$\pm$1.71&0.817$\pm$0.07\\
\Xhline{1.5 pt}
\end{tabular}
\end{table*}
\end{center}

\begin{center}
\begin{table*}[t]
\caption{Quantitative results of ablation studies.}
\label{table2}
\centering
\begin{tabular}{ccccc}
\Xhline{1.5 pt}
Method& SSIM$\uparrow$ & MAE(HU)$\downarrow$ & PSNR(dB)$\uparrow$ & NCC$\uparrow$ \\
\Xhline{0.5 pt}
with Energy-Guided function &\pmb{0.850}$\pm$\pmb{0.03} & \pmb{26.87}$\pm$\pmb{5.90} & \pmb{19.83}$\pm$\pmb{1.39}& \pmb{0.874}$\pm$\pmb{0.04}\\
without Energy-Guided function &0.825$\pm$0.04 & 27.19$\pm$5.96 & 18.87$\pm$1.71&0.817$\pm$0.07\\
\Xhline{1.5 pt}
\end{tabular}
\end{table*}
\end{center}

 We collect CT and CBCT images from 441 patients who undergo chest cancer radiation therapy at West China Hospital, Sichuan University. Each patient has a set of CT images and corresponding five sets of CBCT images, resulting in 2,205 data samples for training. We anonymize all CT and CBCT images to protect personal information. West China Hospital, Sichuan University approves the study. We collect CT images using the Ge Medical System and CBCT images using the Varian Medical System.

The voxel size of CT images is $0.9766 \times 0.9766 \times 3$, with dimensions of $512 \times 512$. Each patient generates 100 to 250 slices of CT images. The voxel size of CBCT images is $0.5112 \times 0.5112 \times 1.99$, with dimensions of $512 \times 512$. Each patient generates 93 slices of CBCT images. We perform rigid registration using the Ants registration framework to ensure consistency between CT and CBCT images in voxel size and the number of slices. This registration aligns the voxel size and the number of slices of CBCT images with their corresponding CT images. As a result, we obtain a total of 98,555 unpaired slices. Among them, 89,000 slices are used as the training set, 437 as the validation set, and 9,118 as the test set.

\subsection{Implementation Details}
We compare our method with CycleGAN\cite{JunYanZhu2017UnpairedIT}, CUT \cite{TaesungPark2020ContrastiveLF}, NEGCUT\cite{WeilunWang2021InstanceWiseHN}, MoNCE\cite{FangnengZhan2023ModulatedCF}, Qs-Attn\cite{XueqiHu2023QSAttnQA}, DDPM\cite{Lyu_Wang_2022}. All models are implemented in Python using the PyTorch framework. The models use the Adam optimizer with $\beta_1$ = 0.5, $\beta_2$ = 0.9, and other parameter settings following\cite{Lyu_Wang_2022}. We evaluate the synthesized performance by Peak Signal to Noise Ratio (PSNR), Structural Similarity (SSIM), Mean Absolute Error (MAE), Normalized Cross-Correlation(NCC) metrics.

\subsection{Results}
We demonstrate the effectiveness of EGDiff in the unsupervised CBCT-based sCT generation on a chest tumor dataset provided by West China Hospital. We compare the results with state-of-the-art non-attention GAN models (CycleGAN, CUT, QS-Attn, MoNCE, and NEGCUT) and conventional diffusion (DDPM) models. Figure~\ref{fig2} shows the sCT generated by EGDiff and other comparative methods. The first, third, and fifth rows display the synthesis results, while the second, fourth, and sixth rows show the enlarged synthesis details. Compared to other methods, the sCT generated by EGDiff is more precise and more anatomically consistent with the plan CT (pCT). The zoomed-in regions show that the aorta and heart synthesized by EGDiff have similar spatial dimensions to pCT. In contrast, other methods generate images with excessive or insufficient structures. CycleGAN, CUT, QS-Attn, MoNCE, and NEGCUT generate excessive tissue, while DDPM generates sCT with missing tissue that does not align with pCT. In terms of the overall structure, due to the incompleteness of CBCT, the sCT generated by other methods perform poorly in unknown regions, exhibiting contour deformations. In contrast, the sCT generated by EGDiff maintains consistency with pCT even in the contour. Table~\ref{table1} provides quantitative results of EGDiff and other methods regarding SSIM, MAE, PSNR, and NCC. EGDiff outperforms other models in all four evaluation metrics, with SSIM, MAE, PSNR, and NCC being 0.850, 26.87 HU, 19.83 dB, and 0.874, respectively, indicating that the sCT generated by EGDiff closely resembles pCT.

Figure~\ref{fig1} visualizes the HU value differences between sCT generated by different methods and pCT. Closer proximity to white indicates more minor differences, while closer proximity to blue indicates more significant differences. Compared to other methods, EGDiff exhibits significantly lower HU value differences with pCT, particularly in complex soft tissue regions and contours.

Figure~\ref{fig3} illustrates the horizontal HU line contours of the sCT images generated by all methods and pCT. The x-axis represents the contour range, and the y-axis represents HU values. The zoomed-in regions in the figure shows that the HU values of sCT generated by EGDiff are closer to pCT, particularly in prominent peaks or troughs. Additionally, EGDiff achieves a Pearson correlation coefficient of 0.950 between the generated sCT images and pCT, surpassing other models. In other words, EGDiff better preserves the high-density and edge structures of pCT.

Figure~\ref{fig4} presents the distribution of evaluation metric values for all test samples under different methods. The sCT generated by EGDiff exhibits a relatively concentrated distribution across the four evaluation metrics. At the same time, other forms show more scattered distributions with more considerable differences between minimum and maximum values. It indicates that EGDiff has better mean values and lower standard deviation.

\subsection{Ablation studies}
We conduct ablative experiments to validate the effectiveness of our energy-guided function in chest tumor dataset. We remove the energy-guided function and perform experiments under the same network architecture and parameters. As shown in Table~\ref{table2}, the experimental results demonstrate the impact of eliminating the energy-guided function. Following its removal, the SSIM decreases from 0.850 to 0.825, the MAE increases from 26.87 HU to 27.19 HU, the PSNR decreases from 19.83 dB to 18.87 dB, and the NCC decreases from 0.874 to 0.817. These results indicate that the energy-guided function enhances the model’s generation capability and produces more realistic sCT that closely resemble pCT.

\section{Discussion}
Our study demonstrates the effectiveness of EGDiff in generating unsupervised CBCT-based sCT images for a chest tumor dataset. We compare EGDiff’s performance with several state-of-the-art non-attention GAN models (CycleGAN, CUT, QS-Attn, MoNCE, and NEGCUT) and conventional diffusion models (DDPM).

Figure~\ref{fig2} visually illustrates that EGDiff produces more precise and anatomically consistent sCT images with the reference pCT compared to other methods. The detailed regions, such as the aorta and heart, exhibit similar spatial dimensions to the pCT in EGDiff’s synthesized images. In contrast, other methods tend to generate images with excessive or insufficient structures. CycleGAN, CUT, QS-Attn, MoNCE, and NEGCUT tend to generate excessive tissue, while DDPM fails to capture certain tissue structures in the pCT. Moreover, other methods struggle to maintain structural consistency in unknown regions, resulting in contour deformations. In contrast, EGDiff maintains consistency even in the contour of the generated sCT images.
Table~\ref{table1} quantitatively evaluates the performance of different models using evaluation metrics such as SSIM, MAE, PSNR, and NCC. EGDiff outperforms other models in all metrics, with values of 0.850 (SSIM), 26.87 HU (MAE), 19.83 dB (PSNR), and 0.874 (NCC). These results indicate that the sCT images generated by EGDiff closely resemble the reference pCT.

Figure~\ref{fig1} also visualizes the HU value differences between the sCT images generated by different methods and the pCT. EGDiff exhibits significantly lower HU value differences with pCT, especially in complex soft tissue regions and contours, indicating its ability to capture better and represent the underlying anatomical structures.

Figure~\ref{fig3} shows the horizontal HU line contours of the sCT images generated by all methods compared to the pCT. EGDiff achieves HU values closer to the pCT, particularly in prominent peaks or troughs. Additionally, EGDiff demonstrates a high Pearson correlation coefficient of 0.950, surpassing the performance of other models. It indicates that EGDiff better preserves the high-density and edge structures present in the pCT.

The distribution of evaluation metric values for all test samples under different methods, as shown in Figure~\ref{fig4}, further supports the superior performance of EGDiff. The sCT images generated by EGDiff exhibit a relatively concentrated distribution across the four evaluation metrics. In contrast, other methods show more scattered distributions with enormous differences between minimum and maximum values. It indicates that EGDiff achieves better mean values and has a lower standard deviation, resulting in more consistent and reliable results.

For our upcoming work, we plan to expand the application scope of EGDiff, based on its foundation in adaptive radiotherapy. In adaptive radiotherapy, we adjust individualized treatment plans according to the patient’s anatomical and physiological conditions to improve the accuracy and effectiveness of treatment. We aim to explore the potential of using generated sCT images for tumor target segmentation. Traditional target segmentation methods need help with complex tumor structures and surrounding tissues. In contrast, high-quality sCT images generated by EGDiff could provide more accurate anatomical information, helping to improve these methods. By using sCT images produced by EGDiff for detailed segmentation, we can precisely determine the contours of the tumor area, offering a more reliable basis for target definition in adaptive radiotherapy.

Secondly, we will investigate the potential of using sCT images in generating adaptive radiotherapy plans. Utilizing the sCT images created by EGDiff, we can obtain more detailed patient anatomy and the position and distribution of the tumor and surrounding tissues. It will provide more accurate input data for generating adaptive radiotherapy plans, which can better consider anatomical changes that might occur during radiotherapy.By combining sCT images created by EGDiff and collected images, we can monitor changes in patient anatomy during radiotherapy in real-time. Based on these changes, we can make adaptive adjustments to ensure the ongoing effectiveness of radiotherapy. This real-time adaptive adjustment can minimize damage to surrounding healthy tissue while providing adequate treatment dose coverage of the tumor area.

In summary, we plan to apply EGDiff to adaptive radiotherapy to enhance the individualization and accuracy of radiotherapy plans, thereby improving treatment outcomes. Using high-quality sCT images generated by EGDiff, we aim to achieve more accurate results in target segmentation and radiotherapy plan generation. It will provide better guidance for clinicians and lead to more effective and safer patient treatments.

\section{Conclusion}
In this study, we propose EGDiff, which utilizes a diffusion model and incorporates an energy-guided function to generate sCT images from CBCT images. We conducted experiments on a chest tumor dataset provided by West China Hospital, Sichuan University. The results demonstrate that our method outperforms the baseline model in four evaluation metrics: SSIM, MAE, PSNR, and NCC. This work can improve the quality of CBCT images and thus enhance the clinical treatment efficiency and effectiveness of adaptive radiotherapy.

\section*{Acknowledgement}
This work was supported by Department of Science and Technology of Sichuan Province (RZHZ2022008) and 1.3.5 project for disciplines of excellence, West China Hospital, Sichuan University(20HXJS040).










\bibliographystyle{cas-model2-names}

\bibliography{cas-refs}



\end{document}